\documentstyle[aps,version2,preprint]{revtex}    
\begin{document}

\draft

\begin{title}
One-Particle Spectral Properties of 1D Mott-Hubbard Insulators
\end{title}

\author{J. M. P. Carmelo$^{1,2}$, J. M. E. Guerra$^{2}$,\\
J. M. B. Lopes dos Santos$^{3}$,
and A. H. Castro Neto$^{4}$}
\begin{instit}
$^{1}$ N.O.R.D.I.T.A., Blegdamsvej 17, DK-2100 Copenhagen \O, Denmark
\end{instit}
\begin{instit}
$^{2}$ Department of Physics, University of \'Evora,
Apartado 94, P-7001 \'Evora Codex, Portugal
\end{instit}
\begin{instit}
$^{3}$ CFP and Departamento de F\'{\i}sica da Faculdade de
  Ci\^encias da Universidade do Porto\\
  Rua do Campo Alegre 687, P-4169-007 Porto, Portugal
\end{instit}
\begin{instit}
$^{4}$ Department of Physics, University of California,
Riverside, California 92521, USA
\end{instit}
\receipt{14 April 1999}

\begin{abstract}
We use an exact {\it holon} and {\it spinon} Landau-liquid
functional which describes the holon - spinon interactions
at all scattering orders, to study correlation functions
of integrable multicomponent many-particle problems
showing both linear and non-linear energy bands. Motivated
by recent photoemission experiments, we consider
specific cases when the dominant non-linear band terms are
quadratic and apply our results to the evaluation of
the half-filling 1D Hubbard model one-particle spectral 
functions beyond conformal-field theory.
\end{abstract}
\renewcommand{\baselinestretch}{1.656}   


\narrowtext

Recent experimental investigations on the one-particle spectral
properties of quasi-one-dimensional (1D) insulators \cite{Kobayashi}
have confirmed that the low-energy physics of these materials is
dominated by {\it holons} and {\it spinons}. The discovery of 
these elementary excitations followed from the diagonalization 
of the 1D Hubbard model \cite{Lieb} by Bethe-ansatz (BA). 
Theoretical studies of the one-particle spectral properties
of this model are mainly numerical \cite{Kobayashi,Hanke,Penc}. 
Analytical studies of the problem either focused on the limit 
of infinite on-site Coulomb interaction (when the 
charge - spin separation occurs at all energies) or to the 
metallic phase away from half filling, where conformal-field 
theory (CFT) has provided important information on correlation 
functions \cite{Penc,Frahm,Carmelo96a}. Unfortunately, CFT 
does not allow the evaluation of one-particle correlation
functions for the half-filling Mott insulator and the study of these
functions at finite values of $U$ remains an open problem of great
physical interest.

In this Letter we combine a Landau-liquid functional which
describes exactly the holon and spinon interactions at all
scattering orders to study
the effects of the non-linearity of the holon and spinon bands
in the asymptotics of the 1D Hubbard model one-particle
correlation functions both at half filling and maximum
magnetization. Our results apply directly to the study of
the 1D insulators and allow for the
description of the unusual spectral properties detected 
in the low-dimensional materials \cite{Kobayashi}.

We consider the 1D Hubbard model in a magnetic field $H$ and
chemical potential $\mu$ which describes $N=N_{\uparrow}+N_{\downarrow}$
interacting electrons and can be written as
$\hat{H}=\hat{T}+U[\hat{D}-{1\over 2}\hat{N}]
+\sum_{\alpha= c,s}\mu^h_{\alpha }2{\hat{S}}^{\alpha}_z$.
We have $\hat{T}=-t\sum_{j,\sigma}[c_{j\sigma}^{\dag}c_{j+1\sigma} + h.
c.]$
the ``kinetic energy'' ($t$ is the transfer integral), 
$\hat{D} = \sum_{j}
\hat{n}_{j,\uparrow}\hat{n}_{j,\downarrow}$
measures the number of doubly occupied sites and
$\mu^h_c=\mu$, $\mu^h_s=\mu_0 H$ ($\mu_0$ is the Bohr magneton)
are the chemical potentials. Moreover, we define the charge
and spin operators
${\hat{S }}^c_z=-{1\over 2}[N_a-\hat{N}]$, ${\hat{S }}^s_z=
-{1\over 2}[{\hat{N}}_{\uparrow}-{\hat{N}}_{\downarrow}]$,
${\hat{N}}=\sum_{\sigma}\hat{N}_{\sigma}$,
${\hat{N}}_{\sigma}=\sum_{j}\hat{n}_{j,\sigma}$ where
$\hat{n}_{j,\sigma}=c_{j\sigma }^{\dagger }c_{j\sigma }$
is the occupation number operator 
($c_{j\sigma}^{\dagger}$ and $c_{j\sigma}$ are electron
operators of spin projection $\sigma $ at site $j=1,...,N_a$). 
We define the charge density
$n=n_{\uparrow}+n_{\downarrow}=N/N_a$ and a spin density
$m=n_{\uparrow}-n_{\downarrow}$ in the intervals $0\leq n\leq 1$
and $0\leq m\leq n$, respectively, where
$n_{\sigma}=N_{\sigma}/N_a$. The relevant Fermi momenta are
$k_{F\sigma }=\pi n_{\sigma}$ and $k_F=\pi n/2$. We use units
such that $a=\hbar=1$, where $a$ is the lattice constant.

In the operational BA representation of Ref. \cite{Carmelo96}
the holons and spinons are the {\it holes} of the $c$ and $s$ 
anticommuting pseudoparticles \cite{Carmelo94}, respectively. 
In this basis the Slater-determinant levels of $\alpha$ 
pseudoparticles, with $\alpha=c,s$, are exact energy 
eigenstates of the many-electron problem and the 
total momentum is additive in the pseudoparticle momenta $q$.
The Hamiltonian contains pseudoparticle interaction
terms, yet integrability implies that all the interactions are of
zero-momentum forward-scattering type. The exact many-electron ground
state (GS) is in this representation described by the
$\alpha$-pseudoparticle
distribution, $N_{\alpha}^0(q)=\Theta (q^{(+1)}_{F\alpha}-q)$ and
$N_{\alpha}^0(q)=\Theta (q-q^{(-1)}_{F\alpha})$ with
$0<q<q^{(+1)}_{\alpha}$ and $q^{(-1)}_{\alpha}<q<0$, respectively,
where $q^{(\pm 1)}_{F\alpha}=\pm q_{F\alpha}+O(1/N_a)$ and
$q_{F\alpha}=\pi N_{\alpha}/N_a$. Here $N_c=N$ and $N_s=N_{\downarrow}$
and hence $q_{Fc}=2k_F$ and $q_{Fs}=k_{F\downarrow}$.
The pseudo-Brillouin zones are $q^{(\pm 1)}_{\alpha}=
\pm q_{\alpha}+O(1/N_a)$ and $q_{\alpha}=\pi N^*_{\alpha}/N_a$
where $N^*_c=N_a$ and $N^*_s=N_{\uparrow}$ and hence
$q_{c}=\pi $ and $q_{s}=k_{F\uparrow}$. Note that at half filling,
$N^h_c=N^*_c-N_c=0$ (zero magnetization, $N^h_s=N^*_s-N_s=0$)
and the GS has no holons (spinons). Importantly,
$N_{\alpha}$ and $N^*_{\alpha}$ are always conserving numbers.
The energy eigenstates only involve
the momentum distribution $N_{\alpha}(q)$ and for states
differing from the GS by a small density of
excited $\alpha $ pseudoparticles the excitation-energy spectrum
can be expanded in the momentum deviations,
$\Delta N_{\alpha}(q)=N_{\alpha}(q)-N_{\alpha}^0(q)$,
what leads to a Landau functional \cite{Carmelo94}, $\Delta E_L =
\sum_{i=1}^{\infty}\Delta E_i$, which up to third order
in $i$ reads

\begin{eqnarray}
\Delta E_L & = & \sum_{\alpha,q}\Delta N_{\alpha}(q)
\epsilon_{\alpha} (q) + {1\over L}
\sum_{\alpha ,\alpha'}\sum_{q,q'}\Delta N_{\alpha}(q)
\Delta N_{\alpha'}(q'){1\over 2}\,f_{\alpha,\alpha'} (q,q') +
\nonumber \\
& + & {1\over L^2}\sum_{\alpha,\alpha',\alpha''}\sum_{q,q',q''}
\Delta N_{\alpha}(q)\Delta N_{\alpha'}(q')
\Delta N_{\alpha''}(q'')
{1\over 6}\, g_{\alpha,\alpha',\alpha''} (q,q',q'') + h.o. \, .
\label{EL}
\end{eqnarray}

All the coefficients of the functional (\ref{EL}) can be exactly
derived from the BA equations \cite{Carmelo94}, yet for orders
$i>2$ the calculations are very lengthy. So far
only the $i=1$ band $\epsilon_{\alpha} (q)$ and $i=2$
$f_{\alpha,\alpha'}$ function were derived \cite{Carmelo94}.
>From the band expressions the velocity $v_{\alpha}(q)=d\epsilon_{\alpha}
(q)/dq$ and the function $a_{\alpha}(q)=d v_{\alpha}(q)/dq$ follow.
The $f_{\alpha,\alpha'}$ function expression only involves the
velocity and the two-pseudoparticle phase shift $\Phi_{\alpha,\alpha'}
(q,q')$ \cite{Carmelo94}. This phase shift plays a central
role in our critical theory -- it is an interaction-dependent
parameter associated with the zero-momentum forward-scattering
collision of the $\alpha$ and $\alpha'$ pseudoparticles of momentum
$q$ and $q'$, respectively. For instance, the phase-shift combinations
$\xi_{\alpha\alpha'}^i=\delta_{\alpha,\alpha'}+\sum_{j=\pm 1}(j)^i
\Phi_{\alpha\alpha'} (q_{F\alpha},jq_{F\alpha'})$ with $i=0,1$
fully determined the $\alpha $ anomalous dimensions of the generalized
critical theory, as we discuss below.

When both the densities of $\alpha $ pseudoparticles, $n_{\alpha}$,
and of $\alpha $ pseudoholes, $n^h_{\alpha}$, are finite the dominant
$\epsilon_{\alpha} (q)$ band terms are linear in the
momentum for low-energy excitations. We emphasize that in this
case only the terms of order $i=1$ and $i=2$ of the functional (\ref{EL})
are relevant for the low-energy physics. Upon linearization
of the $\epsilon_{\alpha} (q)$ bands and replacement of 
$f_{\alpha,\alpha'} (q,q')$ by $f_{\alpha,\alpha'} 
(q_{F\alpha},\pm q_{F\alpha'})$, the spectrum (\ref{EL}) 
leads to the CFT spectrum of Ref. \cite{Frahm}. This 
reveals that our basis is the natural representation for 
the critical-point operator algebra. For these densities CFT 
provides asymptotic correlation functions expressions 
\cite{Frahm}.

The functional (\ref{EL}) is universal for a wide
class of integrable multicomponent models. As we
illustrate below for the 1D Hubbard model, occurs often in these
systems that as the density $n_{\alpha}$ or $n^h_{\alpha}$ becomes
small the dominant low-energy $\alpha$ band term is non linear and
of the general form $\epsilon_{\alpha} (q)\approx\pm ||q|-Q_{\alpha}|^j
/j!{\cal{M}}_{\alpha}$, where $Q_{\alpha}=0$ when 
$n_{\alpha}\rightarrow 0$ and $Q_{\alpha}=q_{\alpha}$ 
when $n^h_{\alpha}\rightarrow 0$,
${\cal{M}}_{\alpha}$ is a positive generalized {\it mass},
and $j$ is a positive integer, $j=1,2,3,4,....,\infty$.
Thus, $j$ characterizes the curvature of the band at the
pseudo-Fermi points, that is, $j=1$ is linear (the usual case
described by CFT), $j=2$ is quadratic and so on ($j=\infty$
describes the dispersionless band).
We use the notation $\{\alpha,j\}$ for a $\alpha$ band
of type $j$. The $c$ and $s$ bands of the present model are 
represented in Figs. 7 and 8, respectively, of Ref. \cite{Carmelo91}. 
For $U>0$ the $c$ band is quadratic around $q=\pm\pi$ and 
$q=0$ and at $m=0$ and $U>0$ the $s$ band is only 
quadratic around $q=0$. In the limit 
of $U\rightarrow\infty$ its width vanishes and it becomes 
flat, whereas the width of the $c$ band is always $4t$. 
The $c$ and $s$ pseudo-Fermi surfaces are at $q=\pm 
2k_F$ and $q=\pm k_{F\downarrow}$, respectively, and at low energy
the bands can be linearized around them outside the quadratic or flat 
regions. Therefore, there occur band changes $\{c,1\}\rightarrow 
\{c,2\}$ as $n_c\rightarrow 0$ and $n^h_c\rightarrow 0$
when $2k_F\rightarrow 0$ and $2k_F\rightarrow \pi$,
respectively, and $\{s,1\}\rightarrow \{s,2\}$
as $n_s\rightarrow 0$ when $k_{F\downarrow}\rightarrow 0$, and 
$\{s,1\}\rightarrow\{s,\infty\}$ as $U\rightarrow\infty$
at $m=0$ and the $s$ band becomes flat ($\epsilon_{s} (q)=0$ implies
$\{s,\infty\}$). For each $\{\alpha ,j\}$ branch the low-energy 
critical spectrum has contributions from terms up to 
order $i=j+1$ of the functional $(\ref{EL})$.

In this Letter we study the effects of non-linearity on the 
asymptotics of correlation functions. In particular,
following recent photoemission data \cite{Kobayashi}, 
we consider the one $\sigma $ electron (or hole) spectral 
function $A_{\sigma}(k,\omega)$, momentum distribution
$N_{\sigma}(k)=\sum_{\omega} A_{\sigma}(k,\omega)$, and 
density of states $D_{\sigma}(\omega)=\sum_k A_{\sigma}(k,\omega)$. 
There are no previous analytical results for these functions 
at finite values of $U$ and (I) for the $n=1$ Mott-Hubbard 
insulator and (II) the fully polarized ferromagnetic ($m=n$) 
initial GS's. At $n=1$ and $m=0$ there are neither holons nor 
spinons in the initial $N$-GS and the final $N-1$ states 
have one holon and one spinon. Importantly, the numerical 
energy dispersions of Ref. \cite{Kobayashi}, which agree with 
the phtoemission data, can be described by our holon and spinon bands 
$\epsilon_{\alpha}(q)$. At large $U$ the right (or left) 
holon line, which we denote by $\omega^R (k)$ [or $\omega^L (k)$], 
of Fig. 3 of the first paper of Ref. \cite{Kobayashi}  
is in the thermodynamic limit slightly $\omega $ shifted 
and refers to $\omega >0$ (or $\omega<0$) and reads 
$\omega^R (k)=2t +\epsilon_c (\pi/2+k)$ 
[or $\omega^L (k)=2t +\epsilon_c (\pi/2-k)$]. It
corresponds in our representation to the creation of 
the spinon at $q=-\pi/2$ (or $q=\pi/2$) and the holon in the 
domain $\vert q\vert\in [\pi/2\, ,\pi]$ (or
$\vert q\vert\in [0\, ,\pi/2]$). We emphasize that the
final $(N-1)$-GS corresponds to $k=\pi/2=k_F$ and
(due to the slight $\omega$ shift) to $\omega =2t$. 
The right holon line is quadratic in $\omega$ around the GS 
energy because of the $j=2$ character of the $c$ band at 
$q\approx\pm\pi$. To illustrate the physical importance of our 
non-linear critical theory, we evaluate below the 
$\omega $ dependence of the $\omega^R (k)$ peak which 
in the $N_a\rightarrow\infty$ limit exists at $k=k_F=\pi/2$, 
between the $k=3\pi/7$ and $k=4\pi/7$ peaks of the figure.
[Below we measure $\omega$ from the GS, {\it i.e.}
$\omega^R (k)=\epsilon_c (\pi/2+k)$.] We also find that
the $n=1$ non-linear effects lead to a non-Luttinger-liquid 
divergent behavior for $D_{\sigma}(\omega)$.

The low-energy critical theory with both $\{\alpha ,1\}$ and 
$j>1$ $\{\alpha ,j\}$ bands (for our model the total number 
of bands is two) requires the use of terms up to the order $i=j+1$ in the
functional (\ref{EL}). For $i=3$ the calculations 
are lengthy and are technically similar to the the ones presented in
the Appendix of Ref. \cite{Carmelo94} for order $i=2$.
Here they involve substitution of the third-order deviation 
expansions in the BA equations of Ref. \cite{Carmelo94}. This 
leads to $g_{\alpha,\alpha',\alpha''}(q,q',q'')=
{\breve{g}}_{\alpha,\alpha',\alpha''}(q,q',q'')+
{\breve{g}}_{\alpha,\alpha',\alpha''}(q',q'',q)+
{\breve{g}}_{\alpha'',\alpha,\alpha'}(q'',q,q')$ where

\begin{eqnarray}
{\breve{g}}_{\alpha,\alpha',\alpha''}(q,q',q'') & = & 4(\pi)^2\{
a_{\alpha} (q) \Phi_{\alpha,\alpha'} (q,q')\Phi_{\alpha,\alpha''} (q,q'')
\nonumber\\
& + & \sum_{j=\pm 1}\sum_{\alpha'''}j {a_{\alpha'''}\over 3}
\Phi_{\alpha''',\alpha}(jq_{F\alpha'''},q)
\Phi_{\alpha''',\alpha'} (jq_{F\alpha'''},q')
\Phi_{\alpha''',\alpha''} (jq_{F\alpha'''},q'')
\nonumber \\
& + & v_{\alpha}(q)[\Phi_{\alpha,\alpha''} (q,q'')
{d\over dq}\Phi_{\alpha ,\alpha'} (q,q')
+ {\cal{A}}_{\alpha,\alpha',\alpha''} (q,q',q'')]
\nonumber\\
& + & \sum_{j=\pm 1}\sum_{\alpha'''}v_{\alpha'''}
\Phi_{\alpha''',\alpha} (jq_{F\alpha'''},q)
{\cal{A}}_{\alpha''',\alpha',\alpha''} (jq_{F\alpha'''},q',q'')\} \, ,
\label{g}
\end{eqnarray}

\begin{eqnarray}
{\cal{A}}_{\alpha,\alpha',\alpha''} (q,q',q'') & = &
\Phi_{\alpha ,\alpha'} (q,q')
{d\over dq}\Phi_{\alpha ,\alpha''} (q,q'')
\nonumber \\
& + & 2\pi\rho_{\alpha}\left(r^{(0)}_{\alpha}(q)\right){d\over dq}
\{[{\Phi_{\alpha',\alpha} (q',q)\Phi_{\alpha',\alpha''} (q',q'')
\over 2\pi\rho_{\alpha'} \left(r^{(0)}_{\alpha'}(q')\right)} +
{\Phi_{\alpha'',\alpha} (q'',q)\Phi_{\alpha'',\alpha'} (q'',q')
\over 2\pi\rho_{\alpha ''} \left(r^{(0)}_{\alpha''}(q'')\right)}]
\nonumber \\
& + & \sum_{j=\pm 1}\sum_{\alpha'''}
j [{\Phi_{\alpha''',\alpha} (jq_{F\alpha'''},q)
\Phi_{\alpha''',\alpha'} (jq_{F\alpha'''},q')
\Phi_{\alpha''',\alpha''} (jq_{F\alpha'''},q'')
\over 2\pi\rho_{\alpha '''} \left(r_{\alpha'''}\right)}]\} \, ,
\label{A}
\end{eqnarray}
$v_{\alpha }\equiv v_{\alpha}(q_{F\alpha })$,
$a_{\alpha }\equiv a_{\alpha}(q_{F\alpha })$,
the function $2\pi\rho_{\alpha}\left(r\right)$ is
such that $1/2\pi\rho_{\alpha}\left(r^{(0)}_{\alpha}(q)\right)=
dr^{(0)}_{\alpha}(q) /dq$, $r^{(0)}_{c}(q)=K^{(0)}(q)$,
$r^{(0)}_{s}(q)=S^{(0)}(q)$, $K^{(0)}(q)$ and
$S^{(0)}(q)$ are the rapidity functions obtained by solving 
Eqs. $(A1)$ and $(A2)$ of Ref. \cite{Carmelo94} for
$N^0(q)$, and $r_{\alpha}=r^{(0)}_{\alpha}(q_{F\alpha})$.

The Landau character of the spectrum (\ref{EL}) implies
that {\it the only} small parameter which defines the
range of validity of the corresponding low-energy critical
theory is the density of excited pseudoparticles. The
associate low-energy processes can produce changes in the
numbers $N_{\alpha \iota}$ of right ($\iota =+1$) and left
($\iota=-1$) pseudoparticles (with $\iota =sgn (q)1=\pm 1$),
which are conserved independently. Equivalently, they can
produce changes in the charge and current numbers $N_{\alpha
}=\sum_{\iota} N_{\alpha \iota}$ and $J_{\alpha }=(1/2)
\sum_{\iota}\iota N_{\alpha \iota}$, respectively. The
above processes can  be classified into three types, (i)
GS - GS transitions associated with
variations $\Delta N_{\alpha }=N_{\alpha }-N_{\alpha }^0$
which only change $q_{F\alpha }$, (ii) finite  momentum
$K=\sum_{\alpha}{\cal{D}}_{\alpha }2q_{F\alpha }$ processes
associated with variations ${\cal{D}}_{\alpha }=J_{\alpha
}-J_{\alpha }^0$, and (iii) a number $N^{\alpha\iota}_{ph}=
0,1,2,...$ of elementary particle - hole processes around
the Fermi point $q^{(\iota)}_{F\alpha}$. In the absence of
transitions (i) we define $q_{F\alpha }$ relatively to the
initial GS and otherwise relatively to the final
GS. Independently of the form of the band $\epsilon_{\alpha } 
(q)$, at low energy the above processes lead
to a momentum of the form $K=k_0+\sum_{\alpha}\Delta P_{\alpha}$
where $\Delta P_{\alpha}={2\pi\over N_a}[\Delta N_{\alpha }
{\cal{D}}_{\alpha } + \sum_{\iota}\iota N_{\alpha\iota}^{ph}]$
and in the present case, $k_0=\sum_{\alpha }{\cal{D}}_{\alpha }
2q_{F\alpha}$. Let us denote by $\sum_{\bar{\alpha}}$,
$\sum_{\breve{\alpha}}$, and $\sum_{\alpha}$ the summations
over linear branches, quadratic (or other $j>2$ non-linear)
branches, and all types of branches, respectively.
The $\{\alpha ,2\}$ band reads $\epsilon_{\alpha} (q)
=\pm ||q|-Q_{\alpha}|^j/2m^*_{\alpha}$, with ${\cal{M}}_{\alpha}
\equiv m^*_{\alpha}$ and $m_{\alpha}^*=1/a_{\alpha} (0)$
and $m_{\alpha}^*=1/a_{\alpha}(q_{\alpha})$ for pseudoparticles
and pseudoholes, respectively. These masses play the same role
as the effective mass of Ref. \cite{McGuire}. For the
Hubbard model $m_{s}^*$ and $m_{c}^{*}$ are given by
Eqs. $(47)$ and $(48)$, respectively, of Ref. 
\cite{Carmelo91}. Introducing in Eqs.
(\ref{EL})-(\ref{A}) the $\{\alpha ,j\}$ band expressions
(with $j=1$ or $j=2$) and the low-energy deviations associated
with all low-energy processes and performing the $q$ summations
leads to the critical spectrum

\begin{equation}
\Delta E = {2\pi\over L}[\sum_{\bar{\alpha}}v_{\alpha }
\Delta^E_{\alpha} + \sum_{\breve{\alpha}}{\breve{v}}_{\alpha}
({\cal{N}}^1_{\alpha})^2] + \sum_{\breve{\alpha}}\{
({\breve{v}}_{\alpha})^3{L\over 6\pi}(m_{\alpha}^*)^2
+ ({2\pi\over L})^2{{\cal{E}}_{\alpha}\over m_{\alpha}^*} +
{2\pi\over L}{\breve{v}}_{\alpha} N_{\alpha}^{ph}\} \, ,
\label{MDE}
\end{equation}
where $\Delta^E_{\alpha}=\sum_{i=0,1}({\cal{N}}^i_{\alpha})^2
+N_{\alpha}^{ph}$, ${\cal{N}}^0_{\alpha}=\sum_{\alpha'}
\xi_{\alpha\alpha'}^0{\Delta N_{\alpha'}\over 2}$,
${\cal{N}}^1_{\alpha}=\sum_{\alpha'}\xi_{\alpha\alpha'}^1
{\cal{D}}_{\alpha'}$, ${\breve{v}}_{\alpha}=
v_{\alpha}\xi_{\alpha}^0$
with ${\breve{v}}_{\alpha}=
q_{F\alpha}/{\breve{m}}_{\alpha}$ and
$v_{\alpha}=q_{F\alpha}/m_{\alpha}^*$
(or ${\breve{v}}_{\alpha}=
[q_{\alpha}-q_{F\alpha}]/{\breve{m}}_{\alpha}$ and
$v_{\alpha}=[q_{\alpha}-q_{F\alpha}]/m_{\alpha}^*$),
$\xi_{\alpha}^0 = {m_{\alpha}^*\over {\breve{m}}_{\alpha}}
= {{\breve{v}}_{\alpha}\over v_{\alpha}} =
{2{\cal{N}}^0_{\alpha}\over \Delta N_{\alpha}}$,
and ${\cal{E}}_{\alpha}$ is independent of
${\breve{v}}_{\alpha}$ and contains no pseudoparticle
interaction parameters. Note that
$q_{F\alpha}=\pi\Delta N_{\alpha}/N_a$
(or $q_{\alpha}-q_{F\alpha}=-\pi\Delta N_{\alpha}/N_a$)
in the limit when $n_{\alpha}=0$ (or $n^h_{\alpha}$=0) for the initial
GS.

When all $\alpha $ bands are linear, the critical-theory expressions
involve the anomalous dimensions, $2\Delta_{\alpha}^{\iota}=
\sum_{i=0,1}({\cal{N}}^i_{\alpha})^2+\iota \Delta N_{\alpha}
{\cal{D}}_{\alpha}+2N_{\alpha,\iota}^{ph}$ (equal to
$(\sum_{i=0,1}\iota^i{\cal{N}}^i_{\alpha})^2+
2N_{\alpha,\iota}^{ph}$ when there is dual symmetry) and
the asymptotics of correlation functions is given by CFT
\cite{Frahm} and reads $\chi_{\vartheta}(x,t) \propto
{\prod_{\alpha,\iota}[e^{-ik_0 x}/
(x-\iota v_{\alpha}t)^{2\Delta^{\iota}_{\alpha}}}]$.
On the other hand, the critical theory associated
with the low-energy spectrum
(\ref{MDE}) is characterized by two remarkable properties: (I)
in spite of its non-linear velocity terms, {\it the same} above
anomalous dimensions are obtained from the following
equation, $2\Delta_{\alpha}^{\iota}=
{L\over 2\pi}[\delta_{\alpha ,{\bar{\alpha}}}\, d\Delta E/dv_{\alpha} +
\delta_{\alpha ,{\breve{\alpha}}}\, d\Delta E/d{\breve{v}}_{\alpha}
+\iota\Delta P_{\alpha}]$. Since their expressions are the
same as in the linear regime, their values are given by taking
the corresponding limits of the densities in the general
expressions -- in the present model the values of the dimensions
$2\Delta_{c}^{\iota}$ (or $2\Delta_{s}^{\iota}$) are in the quadratic
case given by the limit $n\rightarrow 1$ (or $m\rightarrow n$) of the
corresponding dimensions of the linear regime; and
(II) as $n_{\alpha}\rightarrow 0$ or $n^h_{\alpha}\rightarrow 0$
there is an {\it adiabatic} symmetry \cite{note} which implies that
$\xi_{\alpha\alpha'}^0 \, ,\xi_{\alpha'\alpha}^1 \rightarrow
\delta_{\alpha,\alpha'}$ in these limits. (However, if the $\alpha '$
band is linear, $\xi_{\alpha'\alpha}^0$ and $\xi_{\alpha\alpha'}^1$
remain in general interaction dependent.) This symmetry leads to
$\xi_{\alpha}^0\rightarrow 1$, {\it i.e.}
the pseudoparticle interactions which renormalize the quadratic
mass $m_{\alpha}^*$ and associate velocity $v_{\alpha}$ vanish
(and ${\breve{m}}_{\alpha}=m_{\alpha}^*$). For instance,
in the present model
$\xi_{c\alpha}^0\, , \xi_{\alpha c}^1\rightarrow \delta_{c,\alpha}$
as $n\rightarrow 1$ ({\it i.e.}, $n^h_c\rightarrow 0$) and
$\xi_{s\alpha}^0\, , \xi_{\alpha s}^1\rightarrow \delta_{s,\alpha}$
as $m\rightarrow n$ ({\it i.e.}, $n_s\rightarrow 0$).

These symmetries imply that the terms of the critical spectrum 
(\ref{MDE}) which contain no linear velocity terms are of 
non-interacting pseudoparticle character. Following this 
remarkable property we find the following expressions 
for the asymptotics of correlation functions

\begin{eqnarray}
\chi_{\vartheta } (x,t) & \propto &
\prod_{\breve{\alpha},\bar{\alpha} ',\iota'}e^{-ik_0 x}/
[(x-\iota' v_{\alpha '}t)^{2\Delta^{\iota'}_{\alpha'}}
\,(x)^{\sum_{\iota}2\Delta^{\iota}_{\alpha}}] \, ,
\hspace{0.5cm} x \gg \sqrt{2t/m_{\alpha }^*}  \, ,
\hspace{0.5cm} \alpha\in\breve{\alpha}  \nonumber \\
& \propto &
{\prod_{\breve{\alpha},\bar{\alpha}',\iota'} 
1/[(-\iota' v_{\alpha'}t)^{2\Delta^{\iota'}_{\alpha'}}
\,(\sqrt{2t/m_{\alpha }^*})^{\sum_{\iota}2\Delta^{\iota}_{\alpha}}}]
\, , \hspace{1cm} x = 0 \, .
\label{cf}
\end{eqnarray}
These $x$ and $t$ dependences can be understood in the following
way. In the limit of low energy each $\alpha $ excitation branch
corresponds to an independent momentum-energy tensor component and
to one independent Minkowski space with {\it light velocity}
$v_{\alpha}$. For bands with both finite $n_{\alpha}$ and
$n^h_{\alpha}$ densities the velocity $v_{\alpha}$
is also finite and the metric is Lorentzian. However, for
vanishing small values of $n_{\alpha}$ or $n^h_{\alpha}$ the
metric becomes Galilean. In the Minkowshian case the
asymptotic of correlation functions involves the variables
$(x\pm v_{\alpha}t)$ associated with Lorentz
transformations. On the other hand, in the case of Galilean symmetry
space $x$ and time $t$ are transformed independently. This is
consistent with the two asymptotic-expression regimes involving
either $x$ or $t$. Moreover, the only combination of the 
time $t$ and mass $m_{\alpha}^*$ with dimensions 
of $x$ is $const\times\sqrt{t/m_{\alpha}^*}$.
Note that the changes in the asymptotics only
concern the metric whereas the $\alpha$ anomalous dimensions,
whose values depend on $U/t$, $n$, and $m$, remain the
same. Both the CFT asymptotic correlation function expressions and
expressions (\ref{cf}) are limiting cases valid for finite and
vanishing, respectively, values of $n_{\alpha}$ or $n^h_{\alpha}$. As
$n_{\alpha}$ or $n^h_{\alpha}$ is gently increased, we come into a
small-density $\{\alpha, 2\}\rightarrow\{\alpha ,1\}$ band
transition regime which is not described by these asymptotic
expressions.

Our theory {\it does not} describe the case when all bands 
are of $j>1$ type. When some of the bands are of $j>2$ type we 
find expression (\ref{cf}) for $x \gg (t/{\cal{M}}_{\alpha})^{1/j}$,
whereas the $x=0$ expression becomes
$\chi_{\vartheta}(x,t)\propto {\prod_{\breve{\alpha},\bar{\alpha}',\iota'}
1/[(-\iota' v_{\alpha'}t)^{2\Delta^{\iota'}_{\alpha'}}\,
(t/{\cal{M}}_{\alpha})^{{\sum_{\iota}2\Delta^{\iota}_{\alpha}\over j}}}]$.
Here $j$ is meant to be a function of $\alpha$, {\it i.e.}
different $\alpha$ bands may have different $j>1$ values.

Fourier transforms of the above asymptotic expansions provide
correlation-function expressions for values of momentum 
$k$ close to $k_0$ and low values of energy $\omega$ 
measured from the initial GS energy. For $m=0$ and both 
$n=1$ and small finite densities of holes $\delta =(1-n)$ 
and low negative (positive) values of $\omega $ and/or 
values of $k$ close to $k_F$ our generalized theory leads to
$A_{\sigma} (k_F,\omega)\propto |\omega|^{-7/8}$ 
and $N_{\sigma}(k)\propto |k-k_F|^{1/8}$ for the particle
(hole) spectral function and momentum distribution.
(At $n=1$ we define the ground-state energy at the bottom and
the top of the Mott-Hubbard gap \cite{Lieb} for particles and
holes, respectively.) We emphasize that  
$A_{\sigma} (k_F,\omega)\propto |\omega|^{-7/8}$
is (for particles) the above $n=1$ peak of the figure.

On the other hand, the particle (hole)
density of states is given by $D_{\sigma}(\omega)\propto
|\omega|^{-3/16}$ and $D_{\sigma}(\omega)\propto
|\omega |^{1/8}$ for $n=1$ and small finite
$\delta $ values, respectively. However, the latter
expression (also predicted by CFT) is
restricted to frequencies $|\omega |
<E_c=\delta ({2\pi\over L})^2 {5\over 32\, m_c^*}$.
In the limit of $n\rightarrow 1$ this domain shrinks
to a single point and the spectral function diverges as
$D(\omega)\propto |\omega |^{-3/16}$.
In the $m=n$ case we consider creation of one $\downarrow$ 
electron and find for both $m=n$ and for small finite 
densities $n_{\downarrow} $,
$A_{\downarrow} (k_{F\downarrow},\omega)\propto
|\omega|^{-1+{1\over 2}[1-\eta_0]^2}$ and
$N_{\downarrow}(k)\propto 
|k-k_{F\downarrow}|^{{1\over 2}[1-\eta_0]^2}$,
where $\eta_0=(2/\pi)\tan^{-1}([4t\sin (\pi n)]/U)$
and $n<1$, whereas the density of states is given by
$D_{\downarrow}(\omega)\propto
|\omega|^{-{1\over 2}+{1\over 2}[1-\eta_0]^2}$
and $D_{\downarrow}(\omega)\propto
|\omega|^{{1\over 2}[1-\eta_0]^2}$ for $m=n$ and small
finite values of $n_{\downarrow}$, respectively.
The latter expression is restricted to energies
$|\omega | <E_s=\delta ({2\pi\over L})^2 {1\over m_s^*}$.
Again, in the limit of $n_{\downarrow}\rightarrow 0$ this
domain shrinks to a point and the spectral function
behaves as $D_{\downarrow}(\omega)\propto
|\omega|^{-{1\over 2}+{1\over 2}[1-\eta_0]^2}$.
These $D_{\downarrow}(\omega)$ expressions
are not valid for $U\rightarrow\infty$
because then the bands are of $\{c,1\}$ and 
$\{s,\infty\}$ type and instead $D_{\downarrow}(\omega)\propto 
|\omega|^{-{1\over 2}}$. Finally, concerning the 
comparison of our results with previously obtained 
$m=0$ and $U\rightarrow\infty$ expressions,
while our theory does not apply to the $n=1$ case \cite{Carmelo96a}
of bands $\{c,2\}$ and $\{s,\infty\}$,
it provides expressions for the $\{c,1\}$ and $\{s,\infty\}$
case which corresponds to finite values of $n$
and $\delta$. Importantly, our general expressions
lead to the same results as Ref. \cite{Penc}, with the density of
states given by $D_{\sigma}(\omega)\propto
|\omega|^{-3/8}$ and $D_{\sigma}(\omega)\propto
|\omega |^{1/8}$ for $U\rightarrow\infty$ and
small finite values of $4t/U$, respectively
\cite{note1}. Our non-linear critical theory results 
are expected to shed new light on the unusual properties 
of quasi-1D materials.

We thank D. K. Campbell, A. Luther, L. M. Martelo,
and A. W. Sandvik for illuminating  discussions and 
the support of PRAXIS under Grants No. 2/2.1/FIS/302/94 
and BCC/16441/98. A. H. C. N. acknowledges support 
from the Alfred P. Sloan Foundation and the partial 
support provided by an US Department of Energy CULAR 
research grant.



\begin{references}
\bibitem[11]{note}
        In analogy with the adiabatic vanishing of the Fermi-liquid
        quasiparticle interactions, which here is achieved by
        adiabatically changing $n_{\alpha}$ or $n^h_{\alpha}$.
\bibitem[12]{note1}
        By providing the suitable form for the $G(x,t)$ function of 
        page 897 of Ref. \cite{Penc}, our theory explains how the $1/x$
        divergence of the $u_s=0$ integral $\int dt\, G(x=0,t)
        e^{i\omega t}$ is removed.
\end{references}
\end{document}